\documentclass[prd, twocolumn,floats,floatfix,nofootinbib,11pt] {revtex4-1}
\usepackage{graphicx}
\usepackage{dcolumn}
\usepackage{bm}
\usepackage{graphics}
\usepackage{slashed}
\usepackage{amssymb}
\usepackage{natbib}
\usepackage{amsmath}
\usepackage{amsthm}
\usepackage{url}
\usepackage[usenames,dvipsnames,svgnames,table]{xcolor}
\newcommand{\bea}{\begin{eqnarray}}
\newcommand{\ena}{\end{eqnarray}}
\newcommand{\bean}{\begin{eqnarray*}}
\newcommand{\enan}{\end{eqnarray*}}

\begin{document}

 \title{Characteristic analysis for integrable soliton models\\
  on two-dimensional target spaces}
 \author{E. Goulart}
\affiliation{CAPES Foundation, Ministry of Education, Bras\'ilia/DF - Brazil\\
and\\
D.A.M.T.P.,\\  University of Cambridge, U.K.\\egoulart@cbpf.br}
\begin{abstract}
We investigate the evolutionary aspects of some integrable soliton models whose Lagrangians are derived from the pullback of a volume-form to a two-dimensional target space. These models are known to have infinitely many conserved quantities and support various types of exact analytic solutions with nontrivial topology. In particular, we show that, in spite of the fact that they admit nice smooth solutions, wave propagation about these solutions will always be ill-posed. This is related to the fact that the corresponding Euler-Lagrange equations are not of hyperbolic type.
\end{abstract} 
 \maketitle

\section{Introduction}

Nonlinear theories supporting topologically protected solitons appear in a variety of physical situations, from nuclear interactions to condensed matter systems \cite{Manton}. Typically, one starts with a source space-time  $(\textbf{M},g)$ and consider fields $\varphi$ taking values not in a vector space, but rather in a smooth target Riemanniann manifold $(\textbf{N},h)$. The topological character of solutions appears explicitly when we look for static configurations of finite self-energy. Then, hypothesis concerning the behaviour of fields at spatial infinity effectively imply that they can be classified according to some homotopy class. Roughly, this means that solutions are disjoint and cannot be smoothly deformed into each other due to topological reasons. Relevant examples of soliton theories of this type include the $O(k)$ $\sigma$ models \cite{Tataru}, the $SU(2)$ Skyrme model of pions \cite{Skyrme}, and the $\mathbb{S}^{2}$ Faddeev-Niemi model of knotted hopfions \cite{Faddeev}. 

The point of this letter is to shed light into the evolutionary aspects of some integrable soliton models whose Lagrangians are derived from the pullback $\varphi^{*}\epsilon$  of a volume-form $\epsilon$ to a target space (see, for instance, \cite{Adam1}). These models are important for various reasons: i) They generalize to higher space-time dimensions some concepts which proved useful in the context of (1+1)-dimensional field theories such as the zero curvature of Lax-Zakharov-Shabat \cite{Lax}. ii) The corresponding system of quasilinear second order partial differential equations (PDE's) \cite{Courant} is somewhat simpler and often enables analytical treatment of soliton dynamics. iii) They can play a role in the phenomenological description of physically relevant systems such as non-perturbative regimes of QCD and BPS models of baryons at large $N_{c}$ \cite{BPS0,BPS1,BPS2}. Besides - and perhaps even more persuasively - the models are known to have infinitely many conserved quantities and support various types of exact solutions with nontrivial topology. In particular, they predict the existence of static and time-dependent hopfions in $(1+3)$ space-time dimensions and exotic textures for higher dimensions \cite{foot1}. 

The main question we address here is whether the notion of integrability (in the generalised sense of \cite{Adam1} and \cite{Ferreira1}) is compatible with the hyperbolic nature one usually expects in a relativistic field theory \cite{Geroch}. For the sake of simplicity we restrict ourselves to the case of a two-dimensional target space. In particular, we show that, despite the equations of motion admitting infinite exact analytical solutions, their linearised versions are always ill-behaved. In other words, small disturbances around a given solution will, in general, propagate in an undesirable way, signalling to instabilities. We shall see that this result does not depend on the topology of the target manifold but, rather, is due to the specific type of nonlinearities present in the Euler-Lagrange equations. In particular, a detailed analysis reveals that the associated PDE's are not of hyperbolic type.

The organization of this paper is as follows. In section \textbf{II} we review some geometrical aspects of the Lagrangian theory of maps and establish the basic mathematical machinery. Section \textbf{III} deals with maps whose Lagrangians are derived from the pullback of a volume-form to a target space. Here we start with a general framework and then restrict ourselves to the simpler case of a two-dimensional target space. We continue in section \textbf{IV} by investigating the characteristic surfaces of the model and discussing the evolutionary aspects of the linearized equations of motion. We conclude in section \textbf{V} by mentioning a number of possible extensions that could lead to a well behaved theory.

\section{Geometrical Framework} 

We start, briefly, placing the theory in the general geometrical setting of harmonic/wave mappings. Although this construction is hardly employed in the context of solitons, we believe that it is particularly adequate for our purposes and sets the background for future related problems. Our steps here proceed very much in the same way as Eells and Sampson \cite{Eells} and Choquet-Bruhat \cite{Bruhat} (see also \cite{Misner} for some physical motivations). 

\subsection{Kinematics}

Write ($\textbf{M}^{1+m}$,\ $g$) for a $(1+m)$-dimensional space-time with metric signature $(+,-,-,...)$, ($\textbf{N}^{n}$,\ $h$) for a n-dimensional riemannian manifold with metric $h$ and consider smooth maps of the form
\begin{equation}
 \varphi: \textbf{M}^{1+m}\rightarrow \textbf{N}^{n}.
\end{equation} 
If $x^{a}$ $(a=0,...,m)$ and $y^{A}$  $(A=1,...,n)$ denote local coordinates in the base and target, respectively, the map reads $y^{A}=\varphi^{A}(x^{a})$. The \textit{differential} of $\varphi$ at $x\in \textbf{M}^{1+m}$ is, in a sense, the best linear approximation of $\varphi$ near $x$, 
\begin{equation}  
d\varphi_{x}: T_{x}\textbf{M}^{1+m}\rightarrow T_{\varphi(x)}\textbf{N}^{n},
\end{equation}
and can be used to pullback tensors living in $\textbf{N}^{n}$ to $\textbf{M}^{1+m}$. In particular, using the convention $\partial_{a}\varphi^{A}\equiv\varphi^{A}_{\phantom a a}$, the pulled back metric $\varphi^{*}h$ reads
\begin{equation}
L_{ab}(x)\equiv h_{AB}(\varphi(x))\varphi^{A}_{\phantom a a}\varphi^{B}_{\phantom a b}.
\end{equation}
The eigenvalues of $L^{a}_{\phantom a b}$ (which is often called the \textit{strain} for the map \cite{Nick}) at $x\in \textbf{M}^{1+m}$ solve the characteristic polynomial
\begin{equation}
\mbox{det}\big(L^{a}_{\phantom a b}-\lambda\delta^{a}_{\phantom a b}\big)=\sum\limits_{k=0}^{1+m}\sigma_{k}(-\lambda)^{1+m-k}=0,
\end{equation}
where $\sigma_{k}$ is the $k$-th elementary symmetric function of $L^{a}_{\phantom a b}$
\begin{equation}\label{sigmas}
 \sigma_{k}=\frac{1}{k!}\delta^{b_{1}...b_{k}}_{a_{1}...a_{k}}L^{a_{1}}_{\phantom a b_{1}}...L^{a_{k}}_{\phantom a b_{k}},
\end{equation}
with $\delta^{b_{1}...b_{k}}_{a_{1}...a_{k}}$ the generalized Kronecker tensor and $\sigma_{0}\equiv0$. The basic idea is to use the invariants $\sigma_{k}$ as ingredients for a general Lorentz invariant Lagrangian prescription. In particular, defining $\textbf{L}:=L^{a}_{\phantom a b}$,
 \begin{eqnarray*}
&&\sigma_{1}=[\textbf{L}],\quad\quad\sigma_{2}=\big([\textbf{L}]^{2}-[\textbf{L}^{2}]\big)/2,\\
&&\ \sigma_{3}=\big([\textbf{L}]^{3}-3[\textbf{L}][\textbf{L}^{2}]+2[\textbf{L}^{3}]\big)/6,
\end{eqnarray*}
with the brackets $[\ \ ]$ denoting trace operation, for conciseness.

\subsection{Dynamics}

When $m>n$, as is always the case for the integrable theories we shall discuss in the next sections, it follows $0\leq \mbox{rank}(\textbf{L})\leq n$ and $\sigma_{k}=0$ for all $k>n$. In this situation the most general first order action is provided by
\begin{equation}\label{action}
S[\varphi]=\int_{\textbf{M}^{1+m}}\mathcal{L}(\sigma_{1},...,\ \sigma_{n},\varphi)\ dv_{g},
\end{equation}
with $dv_{g}$ the element of volume in $\textbf{M}^{1+m}$. The Euler-Lagrange equations can be written in a compact form if we introduce a linear connection in the associated vector bundle $\textbf{E}=T^{*}\textbf{M}^{1+m}\otimes \varphi^{-1}T\textbf{N}^{n}$. If $u^{A}_{\phantom a a}(x)$ is a smooth cross-section of \textbf{E}, its covariant derivative is written as \cite{Misner}.
\begin{equation}\label{covariant}
\mathfrak{D}_{b}u^{A}_{\phantom a a}=\partial_{b}u^{A}_{\phantom a a}-{}^{(\textbf{M})}\Gamma^{c}_{\phantom a ab}u^{A}_{\phantom a c}+{}^{(\textbf{N})}\Gamma^{A}_{\phantom a BC}u^{B}_{\phantom a a}\varphi^{C}_{\phantom a b}
\end{equation}
where ${}^{(\textbf{M})}\Gamma^{c}_{\phantom a ab}$ and ${}^{(\textbf{N})}\Gamma^{A}_{\phantom a BC}$ are the Christoffel symbols corresponding to the the metrics $g$ and $h$, respectively. Generalization to mixed objects with more than one internal index (A) follows the same route: simply add one connection ${}^{(\textbf{N})}\Gamma^{A}_{\phantom a BC}$ term to `covariantise' each of the internal indices \cite{foot2}. 

With this notation the critical points of (\ref{action}) satisfy the nonlinear equations of motion
\begin{equation}\label{Euler}
\mathfrak{D}^{a}K^{A}_{\phantom a a}=\frac{h^{AB}}{2}\frac{\partial\mathcal{L}}{\partial\varphi^{B}},
\end{equation}
where
\begin{eqnarray}\nonumber
&&K^{A}_{\phantom a a}\equiv \sum\limits_{k=1}^n\frac{\mathcal{L}_{k}}{(k-1)!}\Big[\delta_{a\ a_{1}...a_{k-1}}^{ b\ b_{1}...b_{k-1}}\\\label{K}
&&\quad\quad\quad\quad\quad L^{a_{1}}_{\phantom a b_{1}}...\ L^{a_{k-1}}_{\phantom a b_{k-1}}\Big]\varphi^{A}_{\phantom a b},
\end{eqnarray}
and $\mathcal{L}_{k}\equiv \partial\mathcal{L}/\partial\sigma_{k}$ for conciseness. The pattern here is simple: the \textit{k-th} term inside brackets involves $(k-1)$ powers of the \textit{strain}. Written more explicitly, Eq. (\ref{Euler}) becomes 
\begin{eqnarray*}
&&\frac{1}{\sqrt{-g}}\partial_{a}\Big(\sqrt{-g}\ g^{ab}K^{A}_{\phantom a b}\Big)+\\
&&\quad\quad\quad+{}^{(N)}\Gamma^{A}_{\phantom a BC}K^{B}_{\phantom a a}\partial^{a}\varphi^{C}=\frac{h^{AB}}{2}\frac{\partial\mathcal{L}}{\partial\varphi^{B}},
\end{eqnarray*}
revealing that various types of nonlinearities can be present. Generically, the first term leads to quasi-linear contributions while the others give rise to semi-linear terms. In particular, the r.h.s. plays the role of a potential and can be chosen according to different motivations. Note also that the PDE's are covariant with respect to coordinate re-parametrizations both in $\textbf{M}^{1+m}$ and $\textbf{N}^{n}$.

\section{Pullback of the Volume form and Integrable models} 

\subsection{n-dimensional targets}

A class of nonlinear theories which well fits in the geometrical formalism described so far is one where the Lagrangian consists of a function of the pullback $\varphi^{*}\epsilon$ of the pertinent volume form on a given target $\textbf{N}^{n}$,
\begin{equation}\label{pull}
H_{a_{1}...a_{n}}=\epsilon_{A_{1}...A_{n}}\varphi^{A_{1}}_{\phantom a a_{1}}...\ \varphi^{A_{n}}_{\phantom a a_{n}},
\end{equation}
with $\epsilon_{A_{1}...A{n}}$ representing the totally antisymmetric Levi-Civita tensor of dimension $n$. Specifically, one considers a Lagrangian that is proportional to a smooth function of the square of $\varphi^{*}\epsilon$ i.e. $\mathcal{L}\rightarrow\mathcal{L}(H)$ with $H\equiv H_{a_{1}...a_{n}} H^{a_{1}...a_{n}}$. Interestingly, the associated action is a particular instance of (\ref{action}) as it directly follows from definitions (\ref{sigmas}) and (\ref{pull}) that $H=n!\sigma_{n}$.
 
Thus, for theories living in a space-time $\textbf{M}^{1+m}$, with $m>n$, one considers the family of Lagrangians
\begin{equation}\label{derived}
S[\varphi]_{int}=\int_{\textbf{M}^{1+m}}\mathcal{L}(\sigma_{n})\ dv_{g}.
\end{equation}
The restriction $m>n$ is important since in the case where the number of spatial dimensions $m$ coincides with the dimension of the target $n$ the corresponding static equations are trivial \cite{Adam1}. When (\ref{derived}) is further restricted to the form $\mathcal{L}(\sigma_{n})\propto \sigma_{n}^{q}$, with $q$ a positive real parameter, it is possible to choose $q$ so as to avoid Derrick's scaling arguments and various types of exact solitonic solutions were obtained in the case of spherical targets $N^{n}=\mathbb{S}^{n}$ (see, \cite{Ferreira1} and references therein). These solutions are topological in the sense that they're characterised by the corresponding homotopy group $\pi_{m}(\mathbb{S}^{n})\in\mathbb{Z}$.

Models of this form are integrable in the sense that they posses a generalised zero-curvature representation  and an infinite number of local conservation laws discussed in a series of papers \cite{Adam1}, \cite{Ferreira1}, \cite{Ferreira2}, \cite{Ferreira3}. Roughly, the Noether currents are associated to the huge amount of symmetry implied by the invariance of the theory under volume preserving diffeomorphisms on the target space. Using (\ref{K}) we obtain, after some algebra
\begin{equation}\nonumber
K^{A}_{\phantom a a}\propto\mathcal{L}_{n}\epsilon^{A}_{\phantom a A_{1}...A_{n-1}}H_{a}^{\phantom a a_{1}...a_{n}}\varphi^{A_{1}}_{\phantom a a_{1}}...\varphi^{A_{n-1}}_{\phantom a a_{n-1}}.
\end{equation}
Note that, in this form, both the volume-form $\epsilon$ and its pullback $\varphi^{*}\epsilon$ appear explicitly. Now, from (\ref{covariant}) it directly follows
\begin{equation}
\mathfrak{D}_{a}\epsilon^{A}_{\phantom a A_{1}...A_{n-1}}=0\quad\quad  \mathfrak{D}_{[a}\partial_{b]}\varphi^{A}=0,
\end{equation}
which imply that the equations of motion for the integrable models read
\begin{eqnarray}\nonumber
&&\epsilon^{A}_{\phantom a A_{1}...A_{n-1}}\nabla_{a}\big(\mathcal{L}_{n}H^{a a_{1}...a_{n-1}}\big)\times\\\label{imp}
&&\quad\quad\quad\times\varphi^{A_{1}}_{\phantom a a_{1}}...\varphi^{A_{n-1}}_{\phantom a a_{n-1}}=0,
\end{eqnarray}
where, now, $\nabla^{a}$ denotes ordinary covariant differentiation in $M$.

\subsection{2-dimensional targets}    

From now on we shall focus on integrable models living in a (3+1)-dimensional space-time and taking values in a 2-dimensional target space. Apart from these restrictions, we make no further hypothesis on the geometry/topology of $\textbf{M}^{1+3}$ and $\textbf{N}^{2}$. One first construction of this type was considered by Aratyn, Ferreira and Zimerman (AFZ) with base space $\mathbb{R}^{1+3}$ and target the symmetric space  $SU(2)/U(1)\cong\mathbb{S}^{2}$. Assuming the fractional Lagrangian
\begin{equation}
S[\varphi]_{int}=-\frac{1}{2}\int_{\mathbb{R}^{1+3}}\sigma_{2}^{3/4}\ dv_{g}
\end{equation}
so as to bypass the usual obstacle due to Derrick's scalings arguments,  they show that Eq. (\ref{imp}) is solvable and infinitely many static analytical solutions of toroidal symmetry, classified by a Hopf index $Q_{H}=\pi_{3}(\mathbb{S}^{2})\in\mathbb{Z}$, exist  \cite{AFZ1}, \cite{AFZ2}. A second model, which explores the invariance under the conformal group $SO(4,2)$, is provided by the strongly-coupled model
\begin{equation}
S[\varphi]_{int}=-\frac{1}{2}\int_{\mathbb{R}^{1+3}}\sigma_{2}\ dv_{g},
\end{equation}
introduced by Ferreira in \cite{Ferr} (see also \cite{Shi}). In this case, although the solitons cannot be brought to rest there exist exact nontrivial time-dependent hopfions carrying angular momentum. As stressed in \cite{Ferr}, this is a rare example of an integrable theory in four dimensions and its solitons may have some role in the low energy limit of Yang-Mills theory. 

In principle, the assumption of a flat space-time is not mandatory and more ambitious models using curved source spaces have been considered. In \cite{Carli}, for instance, the authors considered a space-time of the form $\mathbb{S}^{3}\times\mathbb{R}$ and the Poincar\'e disk as a target. The important point here is that, for all of these models, Eq. (\ref{imp}) reduces to
\begin{eqnarray}\label{2dim}
\partial_{a}\left(\sqrt{-g}\ \mathcal{L}_{2}H^{ab}\right)\varphi^{A}_{\phantom a b}=0,
\end{eqnarray}
with $H_{ab}=\epsilon_{AB}\varphi^{A}_{\phantom a a}\varphi^{B}_{\phantom a b}$ and $\epsilon_{AB}$ the area two-form on $\textbf{N}^{2}$. The following three facts hold for Eq. (\ref{2dim}): i) it consists a two-dimensional system of second-order, quasi-linear PDE's, for the map; ii) the connection of the target space ${}^{(N)}\Gamma^{A}_{\phantom a BC}$ does not appear explicitly in the equations, implying that various types of self-interactions do not contribute. iii) if $\varphi^{A}(x)$ is a solution of (\ref{imp}) for a given $h_{AB}$ it is also a solution for $\tilde{h}_{AB}$ if $\mbox{det}(h_{AB})=\mbox{det}(\tilde{h}_{AB})$.

\section{Characteristic Analysis} 
The natural question to ask for equations (\ref{2dim}) is whether they pose an initial value formulation: in other words, we would like to determine the complete map $\varphi:(\textbf{M}^{1+3},g)\rightarrow (\textbf{N}^{2},h)$ from its values $\varphi|_{\Sigma}$ and first derivatives $\partial\varphi|_{\Sigma}$ restricted to a space-like sub-manifold $\Sigma\subset \textbf{M}$. For well-posedness to hold, it is crucial that small disturbances (or, equivalently, discontinuities) in the initial data propagate in a controlled/predictable way with a definite finite velocity of propagation. Roughly speaking, this means that the associated characteristic surfaces about a background solution have to be described by algebraic varieties with the topology of convex cones for each point of $\textbf{M}^{1+3}$ (see \cite{GWG} and \cite{GWG1} for a similar analysis in the contexts of the Baby-Skyrme and Faddeev models ). 

In order to clarify this point, it is convenient to rewrite Eq. (\ref{2dim}) as
\begin{equation}\label{JA}
M^{ab}_{\phantom a\phantom a  AB}(\varphi,\partial\varphi)\ \partial_{a}\partial_{b}\varphi^{B}+J_{A}(\varphi,\partial\varphi)=0,
\end{equation}
where $J_{A}(\varphi,\partial\varphi)$ stands for semilinear terms in $\varphi$ (whose explicit form is unnecessary for our discussion) and $M^{ab}_{\phantom a\phantom a  AB}$ is the so-called principal part of the system. Here, $M^{[ab]}_{\phantom a\phantom a  AB}=M^{ab}_{\phantom a\phantom a  [AB]}=0$, with brackets denoting antisymmetrization. A closer inspection of Eq. (\ref{2dim}) gives 
\begin{eqnarray}\nonumber
M^{ab}_{\phantom a\phantom a  AB}&=& \big(\epsilon_{AP}\epsilon_{BQ}-\xi N_{AP}N_{BQ}\big)\partial^{a}\varphi^{(P}\partial^{b}\varphi^{Q)}-\\\label{princpart}
 &&\quad\quad\quad-g^{ab}N_{AB},
\end{eqnarray}
with $N_{AB}\equiv\epsilon_{AP}\epsilon_{BQ}\partial_{c}\varphi^{P}\partial^{c}\varphi^{Q}$ and $\xi\equiv\ 2\mathcal{L}_{22}/\mathcal{L}_{2}$. As it is well known, the principal part almost completely controls the qualitative behaviour of solutions of a PDE \cite{foot3}. In particular, it determines the evolution of linearized waves about some smooth background solution $\varphi_{0}(x)$ and gives rise to the causal structure of the theory on top of this solution. Although some progress has been made in this direction for the Skyrme and related models (see, for example, Wong's paper \cite{Wong}), to the best of our knowledge no other authors have considered the detailed structure of (\ref{JA}) in the case of integrable soliton models.

Physically, the characteristics can be identified with the infinite-momentum limit of the \textit{eikonal} approximation (or, equivalently, with the surfaces of discontinuity obtained via Hadamard's method \cite{Perlick}). Briefly, they can be obtained as follows: recall that, for a covector $k_{a}\in T^{*}_{x}\textbf{M}$, the principal symbol is given by the contraction (see, for instance, \cite{Viss})
\begin{equation}
M_{AB}(k)=M^{ab}_{\phantom a\phantom a AB}k_{a}k_{b}.
\end{equation}
In our case, we have
\begin{eqnarray}\nonumber
M_{AB}(k)&=& |\ell|_{h}^{2}h_{AB}-|k|_{g}^{2}N_{AB}-\ell_{A}\ell_{B}-\\\label{MAB}
&&\quad\quad -\xi N_{AP}N_{BQ}\ell^{P}\ell^{Q}   \end{eqnarray}
with $\ell^{P}=\partial^{a}\phi^{P}k_{a}$ and $|\ell|_{h}^{2}$, $|k |_{g}^{2}$ the usual norms with respect to the corresponding metrics.  
Now, If $\varphi_{0}(x)$ is a smooth solution of (\ref{JA}), the determinant $|M_{AB}(k)|$ is a real valued function of $x$ and $k$ and is called the characteristic polynomial $\mathcal{P}(x,k)$ around the solution $\varphi_{0}(x)$. The set
\begin{equation}
N_{x}=\{k\in T_{x}^{*}\textbf{M}\ \big{|}\ \mathcal{P}(x,k)=0,k\neq 0\}
\end{equation}
is called the characteristic set and consists of the locus of normals to the characteristic surfaces at $x$. As Eq. (\ref{MAB}) is given by a $2\times 2$ matrix, the characteristic polynomial reduces to the form
\begin{equation}
\mathcal{P}(x,k)=G^{abcd}k_{a}k_{b}k_{c}k_{d}
\end{equation} 
with $G^{abcd}$ a completely symmetric tensor density depending on $\varphi_{0}(x)$. Thus, the wave normals are determined by the vanishing sets of a multivariate polynomial of fourth order in $k_{a}$ in the cotangent space. The resulting algebraic variety changes from point to point in a way completely prescribed by the background solution and the nonlinearities of the given integrable model. For such theories (as for quasi-linear theories in general) wave velocities are not given a priori, but change as functions of initial data, directions of propagation and wave polarization \cite{Taniuti}.
  
It is quite common in nonlinear field theories described by second-order PDE's that the characteristic polynomial factorizes into simpler irreducible polynomials. This phenomena happens, for instance, in the context of nonlinear electrodynamics \cite{Plebanski}, Born-Infeld theories \cite{Stringy}, optics inside media \cite{Perlick} and Lovelock theories of gravity \cite{Reall}. The integrable theory of maps into a two-dimensional surface is not an exception. Indeed, for the matrix (\ref{MAB}) a tedious, but straightforward, calculation reveals that the characteristic equation always factorizes into a product of quadratic terms, i.e. 
\begin{equation}\label{z}
\mathcal{P}(x,k)=h\mathcal{P}_{1}(x,k)\mathcal{P}_{2}(x,k)=0,
\end{equation}
with $h\equiv \mbox{det}(h_{AB})$,
\begin{equation}\nonumber
\mathcal{P}_{1}(x,k)\equiv G^{ab}_{(1)}k_{a}k_{b},\quad \mathcal{P}_{2}(x,k)\equiv G^{cd}_{(2)}k_{c}k_{d},
\end{equation}
and
\begin{eqnarray}\label{quantity1}
&&G^{ab}_{(1)}\equiv\sigma_{2}g^{ab}+H^{a}_{\phantom a c}H^{cb},\\\label{quantity2}
&&G^{ab}_{(2)}\equiv g^{ab}-\xi H^{a}_{\phantom a c}H^{cb}.
\end{eqnarray} 
In order to derive these relations we used the fact that, for a 2-dimensional target and a four dimensional base, $\sigma_{3}=\sigma_{4}=0$. In this case the Cayley-Hamilton theorem then implies the relation $\textbf{L}^{4}=\sigma_{1}\textbf{L}^{3}-\sigma_{2}\textbf{L}^{2}$ which is important for the factorization \cite{foot4}.

At first sight it appears that the variety of wave normals is given by a product of quadratic Lorentzian cones in $T_{x}^{*}M$. What about the characteristic surfaces themselves? The theory of PDE's proceeds by showing that they are generated by bi-characteristic rays $x^{a}(\lambda)$. In the case where $\mathcal{P}$ factorizes, then we must use each $\mathcal{P}_{i}$ (i=1,2) instead of $\mathcal{P}$ in defining the characteristics. Consequently, the bi-characteristic rays $\{x^{a}(\lambda)\}$ are solutions of the canonical equations
\begin{equation}\nonumber
\dot{x}^{a}=\frac{\partial \mathcal{P}_{i}}{\partial k_{a}}\quad\quad\dot{k}_{a}=-\frac{\partial \mathcal{P}_{i}}{\partial x^{a}}\quad\quad (i=1,2),
\end{equation}
where the dot means derivative with respect to the parameter $\lambda$. The idea here is to use the first equation to solve for $k_{a}$ in terms of $\dot{x}^{a}$ and use the second to obtain a second order equation for $x^{a}(\lambda)$, as is common in Hamiltonian mechanics/optics. It is clear that this can only be achieved if both the quadratic forms $G^{ab}_{(i)}$ (which depend on the background) are nondegenerate.

Unfortunately, this is not true for the expressions (\ref{quantity1}) and (\ref{quantity2}) because one of the quadratic forms is singular, independently of the Lagrangian. Indeed, for a generic antisymetric tensor $H_{ab}$ living in $(1+3)$ dimensions and an arbitrary function $f$, the following holds
\begin{equation}\nonumber
\big{|}\delta^{a}_{\phantom a b}+f H^{a}_{\phantom a c}H^{c}_{\phantom a b}\big{|}=\mathcal{U}^{2},
\end{equation}\\
with
\begin{equation}\nonumber
\mathcal{U}\equiv 1-\frac{f}{2}H_{ab}H^{ab}-\frac{f^{2}}{16}(H_{ab}\stackrel{\ast}{H^{ab}})^{2}
\end{equation}
and $\stackrel{\ast}{H^{ab}}=\frac{1}{2}\eta^{abcd}H_{cd}$ the dual of $H^{ab}$. Note, however, that, if $H_{ab}$ is the pullback of a volume form, we have
\begin{eqnarray*}
H_{ab}\stackrel{\ast}{H^{ab}}=\frac{1}{2}\eta^{abcd}(L_{ac}L_{bd}-L_{ad}L_{bc})=0,
\end{eqnarray*}
meaning that the last term in $\mathcal{U}$ always vanishes for the integrable models, independently of the background solution. Moreover, recalling that $H^{ab}H_{ab}=2\sigma_{2}$, one obtains 
\begin{equation}\nonumber
|G^{ab}_{(1)}|=0,\quad\quad |G^{ab}_{(2)}|=g^{-1}(1+\xi\sigma_{2}),
\end{equation}
which implies that the characteristic surfaces are not given by a product of Lorentzian cones. Rather, one of them is given in terms of the zeros of a singular quadratic form in $T^{*}_{x}\textbf{M}^{1+3}$. Thus, we conclude that, for all possible Lagrangians depending only on $\sigma_{2}$, the corresponding Euler-Lagrange equations are not of hyperbolic type. Therefore, for these nonlinear field theories, integrability is not compatible with hyperbolicity. As a consequence, in spite of the fact that the models on two-dimensional targets admit infinite sets of exact analytical solutions, their linearized versions are always problematic. In other words, one does not expect that arbitrary initial data will launch a nice solution, even locally in time.

\section{Concluding remarks}

Let us conclude by mentioning a number of possible issues, future directions and open questions. First of all we stress that, for the integrable models taking values on a bi-dimensional target, the associated causal structure is always ill-behaved. In other words, the characteristic surfaces do not have the topology of convex cones, as expected for a hyperbolic theory. This is potentially worrying since small disturbances about any (possibly analytic) smooth solution will not propagate in a well behaved manner in spacetime. What is more, our results do not depend on the specific choice of the Lagrangian neither on the topology/geometry of the target space. Rather, they indicate that the very notion of integrability in the lines of \cite{Adam1} and \cite{Ferreira1} is not compatible with hyperbolicity (well-posedness), at least for a 2-dimensional target space. Physically, this result is somehow expected since these theories are effective and often need corrections.

One may wonder whether slight modifications of the theory could lead to a hyperbolic theory, thus entailing a well-posed Cauchy problem. It first comes to mind the addition of a potential term in the Lagrangian, $\mathcal{U}(\varphi)$. Unfortunatelly, a closer inspection of Eq. (\ref{Euler}) reveals that a potential do not contribute to the principal part of the former PDEs, implying that it is not able to cure the degeneracy of the characteristic surfaces. Another possibility would be to consider general Lagrangians of the form
\begin{equation}
\mathcal{L}=c_{1}\sigma_{1}+c_{2}\sigma_{2}-\mathcal{U}(\varphi)
\end{equation}
where $c_{1}<<c_{2}$, i.e. the Dirichlet term should enter as a rather small addition. Formally, this choice is equivalent to the Faddeev-Niemi model with a potential and admits nice characteristics for sufficiently slowly varying background solutions \cite{GWG1}. However, this model is not integrable and it can be considerably difficult to obtain exact solutions. Hence, given the fact the integrable models are physically compelling and mathematically simple, it would be extremely illuminating to investigate more closely the interconnections between integrability and hyperbolicity.

An extension worth of future investigation is the hyperbolicity of the BPS Skyrme model \cite{BPS0,BPS1,BPS2}. As is well known, the model is based on a Skyrme-type low energy effective action which does have a Bogomolny bound and exact Bogomolny solutions.  It qualitatively reproduces the main features of the liquid droplet model of nuclei and provides quite accurate binding energies of the most abundant higher nuclei (after taking into account the semiclassical rotational and iso-rotational corrections as well as the Coulomb interaction and a small isospin breaking). As the sextic term in its Lagrangian has connections with the pull-back of a volume form to a three-dimensional target space, we expect that similar results presented here would equally hold. This is reinforced by a theorem by Wong (see \cite{Wong}, Theorem 9), which implies that the term $\sigma_{1}$ in the Lagrangian always introduce a factor that is regularly hyperbolic. This in fact has a stabilizing effect on the hyperbolicity of the field theory. The inclusion of this term leads us to the so called near-BPS Skyrme model, which has a chance to be a correct low energy, solitonic model of QCD and nuclear matter. We shall analyse the evolutionary aspects of this model in a forthcomming communication. As a final remark, we note that the BPS limit realizes an important idealization of nuclear matter where i) the classical binding energies are zero ii) the energ-momentum tensor characterizes a perfect fluid \cite{fluid0}. Indeed, the action of the BPS Skyrme model is equivalent to the action of a field theoretic description of perfect fluids in an Eulerian formulation \cite{fluid1,fluid2,fluid3}. Perhaps, the degeneracy of the characteristic surfaces is related to the absence of dissipation in the above model or to the violation of some of the energy conditions. Up to this moment we have no other hints about this unexpected behaviour. It would be extremely interesting to have any deeper insight into this problem.

\section{Acknoweledgement} E. Goulart would like to thank the referee for his/her useful comments and CAPES - Brazil proc. 2383136 for financial support.


\begin{thebibliography}{50}
\bibitem{foot1} For more details, we address the  reader to \cite{Ferreira1}, where the authors give us a thorough analysis of integrability using connections on loop spaces and discuss many possible applications.
\bibitem{foot2} In particular, there follow the compatibility equations $\mathfrak{D}_{a}g_{bc}=0$ and $\mathfrak{D}_{a}h_{BC}=0$, which means that contractions with metrics commute with derivations.
\bibitem{foot3} Note that, the only function in $\Sigma\subset\textbf{M}$ we do not know is $\partial^{2}_{0}\varphi^{A}$. We are supposed to use (\ref{JA}) in order to find these derivatives in terms of the data. This can only be done if the matrix $M^{00}_{\phantom a\phantom a AB}$ is invertible in $\Sigma$.
\bibitem{foot4} It is worth mentioning that expressions of the form (\ref{quantity1}) and (\ref{quantity2}) appear also in nonlinear theories of electrodynamics, such as Euler-Heisenberg and Born-Infeld.
\end{thebibliography}
\end{document}